\documentclass[12pt,fleqn]{article}    
\usepackage{umlaute,latexsym}
\usepackage{epsfig}
\title{Baryon Wave Functions in Covariant  Relativistic Quark Models
\thanks{Supported in part by the Kernforschungszentrum KFZ Juelich,
Germany,  mdillig@theorie3.physik.uni-erlangen.de}} 

\author{M. Dillig \\ 
Institute for Theoretical Physics, University of Erlangen-Nuernberg \\
D-91058 Erlangen, Germany }

\begin{document}
\baselineskip 18pt
\begin{titlepage}

\maketitle

\begin{abstract}
We derive covariant baryon wave functions for arbitrary Lorentz
boosts. Modeling baryons as quark-diquark systems, we reduce their
manifestly covariant Bethe-Salpeter equation to a covariant
3-dimensional form by projecting on the relative quark-diquark
energy. Guided by a phenomenological multigluon exchange
representation of a covariant confining kernel, we derive for
practical applications explicit solutions for harmonic confinement and
for the MIT Bag Model. We briefly comment on the interplay of boosts and
center-of-mass corrections in relativistic quark models. 
\end{abstract}

\end{titlepage}
\newpage
\setcounter{page}{2}
One of the major goals of modern electron and hadron accelerators is
the investigation of the internal structure of hadrons, in particular
of baryons: detailed information is extracted from scattering
experiments at large momentum transfers of typically 1 GeV/c and
beyond. The corresponding form factors map out the various internal
(generalized) charge distributions and provide stringent information
on the underlying quark and gluon degrees of freedom. Presently
various experiments are ongoing with electron or photon beams at
MAMI,ELSA, MIT, JLAB und DESY (1) and with proton beams at COSY and
CELSIUS (2) and other labs. 
\vskip 0.2cm
Practical calculations of form factors suffer in general from the
pertinent problem of center-of-mass (CM) corrections for the many-body
problem and from drastic effects from Lorentz contraction at
increasing momentum transfers. While for the  CM corrections various
recipes have been developed and applied in practical calculations
(3-6), less progress has been achieved in the formulation of covariant
baryon wave functions suitable for practical calculations (7-24). A
possible alternative, the evaluation of formfactors on the light cone,
where Lorentz boosts are completely kinematical, has so far entered
only selectively in practical applications at low scattering energies,
beyond that such an approach suffers from other decreases, such as the
loss of strict rotational invariance (25). As in general the
construction of  boosted, Lorentz contracted wavefunctions is nearly as
complicated as the solution of the full problem, in most practical
applications ad hoc and purely kinematical prescriptions for the
rescaling of the coordinate along the direction of the momentum
transfer are applied (examples are given ref. (26-27)). Thus, specific
questions, as the dependence of Lorentz corrections on the confining
kernel  in quark models, are not addressed. In addition, to hopefully
minimize the influence of Lorentz contractions formfactors are in
general evaluated in the Breit frame, though experimentally they are
measured in the lab system. 
\vskip 0.2cm
In this note we formulate an economical model for covariant baryon
wave functions, which leads to results suitable for practical
applications. As it  our main goal to end up with analytical formulae,
we model the baryon - in the following we use the word proton, though
our approach is fairly general - as a quark-diquark system and restict
ourselves, without any loss of generality, to spin-isospin scalar
diquarks (28). 
\vskip 0.2cm
Our starting point is the manifestly covariant 4-dimensional
Bethe-Salpeter equation (29) 
\begin{equation}
\Gamma = K \, G\, \Gamma \quad \mbox{and} \quad \Psi = G \, \Gamma
\end{equation}
with the vertex function and the Bethe-Salpeter amplitude $ \Gamma $
and $\Psi $, respectively, and the interaction kernel K. In the
two-body Greens function for the quark with mass m and the diquark
with mass m* we fix the relative energy dependence from the covariant
projection on the diquark (30) 
\begin{equation}
G (P, q) = \frac{q\!\!\!/ + m}{q^{2} - m ^{2} - i \epsilon} \; i \pi \delta _{+} 
\left ( ( P - q)^{2} - m^{*2} \right )
\end{equation}
which  results up to $0 \left ( \frac{q^{2}}{2M} \right ) $ in the
single particle Dirac equation for the quark for systems with
arbitrary  overall 4-momenta $ Q = ( E(P) = \sqrt{P^{2} + M^{2}}\, ,
0, 0, P ) $  
\begin{eqnarray}
\left ( \frac{M}{E(P)} \right.  \left. \left ( \epsilon + \frac{P}{M}
\, q_{z} - \frac{{\bf q}^{2}}{2 M} \; \right ) -
(\mbox{\boldmath$\alpha$} {\bf q} + \beta m) \right ) \; \varphi (Q,
{\bf q}) & = & \nonumber\\ 
&& \hskip -4cm 
\frac{1}{E(P)} \int K (Q, {\bf q},{\bf k}) \varphi (Q, {\bf k}) d {\bf k}
\end{eqnarray}
\vskip 0.2cm
Without any details we add a brief comment on the CM corrections in
our model: evidently there is a direct coupling between the internal
and external momenta {\bf q} and {\bf P}, or equivalently,
between boosts and the CM motion. In the rest
system the leading center-of-mass corrections are absorbed for $
\epsilon = m + \epsilon _{b}$, where $ \epsilon _{b} $ is the binding
energy of the quark in 
\begin{equation}
\left ( \frac{q^{2}}{2 \mu} + \epsilon _{b} - V_{n} (r) \right )
\varphi ({\bf r}) = 0 
\end{equation}
with the reduced mass $1 /\mu = 1 /m + 1 /(m + m^{*}) $ for an
arbitrary quark potential $ V_{n} (r)$ (a detailed discussion of CM
corrections are presented in a separate paper). 
\vskip 0.2cm
The decisive step for a practical model is the formulation of a
covariant  interaction kernel in eq.~(4). As the dynamics of the quark
- quark interactions, particularly the microscopic nature of the
confinement, lacks an understanding on the fundamental level of QCD,
all current models in practical calculations rely on phenomenological
formulations of the interaction kernel. Being unable to do better, we
proceed here along similar lines: we assume that the interaction
kernel can be presented as a superposition of appropriately weigthed
gluon exchange contributions; quantitative parameters can be
extracted in comparison with studies to baryon spectroscopy, decay
rates or form factors (31). 
Thus we start from the general kernel
\begin{equation}
K(P, q,k) = \sum _{n} \; \frac{k_{n} (P)}{((q - k)^{2} - m ^{2} + i
\epsilon )^{n+1}} 
\end{equation}
for arbitrary powers of n (which reflect different parametrizations of
the confining kernel) (eq. (5) contains the linear confinement in the
Cornell potential (32)). Upon projecting out the relative energy
dependence this yields the covariant, 3-dimensional kernel 
\begin{equation}
K_{n} (P, q, k) \propto \;_{ {\lim \atop \mu \to 0}} \;
\left(\frac{d}{d \mu^{2}} \right )^n \; \frac{1}{\lambda ^{2} (P) q
_{z} ^{2} + {\bf q}^{2}_{\bot} + \mu ^{2} - i \epsilon} 
\end{equation}
with the "quenching parameter"
\begin{equation}
\lambda (P) = M / \sqrt{M^{2} + P^{2} } \, = M / E(P)
\end{equation}
where we introduced the mass scale $\mu $ (to regularize the Fourier
transform to coordinate space). Already simple power counting signals,
that a kernel with the power n leads to confinement with $\sim
r^{2n}$. Upon performing the corresponding Fourier transform to
coordinate space and performing the limit  
$ \mu \to 0 $  we find 
\begin{equation}
K_{n} (P,q) \to  (1+\beta)/2\,  V_{n} (\sqrt{(z/\lambda(P))^{2} +
\mbox{\boldmath$\rho$} ^{2})} 
\end{equation}
where we introduced for convenience 
the particular Dirac structure of the kernel to
facilitate the evaluation of the resulting Dirac equation. 
Eliminating the small component in eq. (3) with the kernel from
eq. (6) and upon dropping CM corrections and $ \epsilon ^{2}_{b} $
terms for compactness, we end up with the Schroedinger type equation
for the large component of the Dirac equation 
\begin{eqnarray}
& & \ (2m\epsilon_{b} -\lambda^{2}( q_{z}- \frac{P}{M} m)^2 ) - {\bf{q}
 ^{2}_{\bot}} -  
 \nonumber\\
&  & \qquad \qquad -\, V_{n} \left ( \sqrt{(z /\lambda (P))^{2} + 
\mbox{\boldmath$\rho$} ^{2} ) } / R  \right ) \; u (z, \bf{\rho} ) = 0
\end{eqnarray}
(with the typical length scale R; in the rest system the equation
above reduces to the standard spherical Schrödinger type equation for
a particle with mass m). The final equation defines with its
connection to the small component by a simple differentation the full
relativistic covariant quark - diquark wave function for arbitrary
Lorentz systems. In the equation above we see the shortcoming from the
phenomenological nature of the interaction kernel: we absorb the
explicit $\epsilon$ and P dependence of the kernel in the definition
of the energy scale $V_{n} $ for the confining force; including an
explicit P dependence in $ V_{n}$ would require a detailed knowledge
of its microscopic origin. 
\vskip 0.2cm
Approximate or numerical solutions for eq. (9) can be obtained for
different confining szenarios (a more detailed investigation, such as
also of the popular linear (heavy quark) confinement (33), is
presented elsewhere). Here we enter only briefly into two szenarios,
which allow a rigorous analytic solution for arbitrary systems:
i. e. harmonic confinement and bag models in the limit   
$ n \to \infty $ in eq(6).
\vskip 0.2cm

\begin{itemize}
\item[-] Harmonic confinement: \\
With the harmonic kernel defined as (34)
\begin{eqnarray}
K(P,q,k) & = & - 12/\pi _{{\lim \atop  (\mu \to 0)}} \left ( ( d/d
\mu^{2})^{2} (\mu / 2 + (d/d \mu ^{2}) \mu^{3} /3 \right ) \nonumber\\ 
& \to & - \left ( (1/\lambda (P))^{2} (d/d q_{z})^{2} + (d/d{\bf
q}^{2}_{\bot}) \right ) \delta (q_{z}-k_{z}) \delta ({\bf q}_{\bot} -
{\bf k}_{\bot}) \, , 
\end{eqnarray}

the solutions for arbitrary excitations of the baryon are
easily obtained in momentum space. After a redefinition of the
longitudinal momentum
and upon separating the longitudinal and the perpendicular component,
the general solution is given by a product of confluent hypergeometric
functions (35). Here we focus only on the nucleon as the quark -
diquark ground state and obtain explicitly  
\begin{equation}
u(q_{z}, {\bf q}_{\bot}) = N \; e^{- \frac{a^{2}}{2} \;\left ( \lambda
^{2} (q_{z} - \frac{m}{M} P \right ) ^{2} + {\bf q}_{\bot} )^{2}} 
\end{equation}
with the oscillator parameter $ a^{2} = \frac{2}{\sqrt{V_{c}}} $, with
$ V_{c}\propto 1/R^{4} $ 
being the confinement strength and with the ground state energy 
\begin{equation}
\epsilon_{b} (P) \cong (1+ \frac{ P^{2}}{M^{2}}) \cdot \frac{\sqrt{V_{c}}}{m} 
\end{equation}
As expected the standard solution for the spherical harmonic
oscillator is recovered in the rest system, i.e. for P=0 and $\lambda
$(0)=1.As the characteristic result we find a quenching of the
effective P-dependent with size parameter 
\begin{equation}
a ^{2}(P) ^{2} =  (\lambda (P)a )^{2} = \frac{M^{2}}{P^{2} + M^{2}} \; a^{2} 
\end{equation}
which leads to Lorentz quenching in coordinate space along the z -
axis and thus to a significant increase of the longitudinal high
momentum components with increasing P (Fig.1(a,b)); 
\vskip 0.2cm

\item[-]  Bag Model:\\
As mentioned above we generate the Bag from the transition $ n \to
\infty $ in the power of gluon-exchange kernel. As we are unable to
present an analytical solution for arbitrary n (a closed solution for
the z-component exists only in the limit of vanishing binding $
\epsilon _{b} \to 0 $ (35)) we first perform the limit $ n \to \infty
$ and then solve the equation 
\begin{equation}
\left (\epsilon _{b} + \frac{P}{M} \; q _{z} + m - ( \alpha \, {\bf q}
+ \beta m) \right ) \quad u(z, \mbox{\boldmath$\rho$}) = 0 
\end{equation}
with the standard  MIT boundary condition for the large and small
components at $ z = \lambda (P) R$ for the bag radius R. For the large
component the ground state solution can be represented as 
\begin{equation}
u (z, \mbox{\boldmath$\rho$}) =
 N \cos (\frac{k_{z}}{\lambda}\, z) \,\, J_{0} (k_{\bot}
\mbox {\boldmath$\rho$})
\end{equation}
where the $\lambda$ dependence of the z-component again reflects the
quenching of the bag (The extension to excited baryon states again is
straightforward). The quenching fo the bag along the boost momentum is
also reflected in the boundary condition
 $z=\lambda R$ for $ \mbox{\boldmath$\rho$}  =0$,
which for the deformed bag
can be solved only numerically (36). Characteristic results for
3 different boost momenta are presented for the large and small
component of the bag ground state solution in Fig. 2. 
\vskip 0.2cm
Comparing our findings with current more phenomenological recipes we
find that a general and simple extension of the parametrization of the
spherical wave functions and momentum distributions in the rest system
to a boosted system, by rescaling the size parameter of the system,
but keeping otherwise the spherical character of the solutions, is
certainly very unsatisfactory and breaks down completely for boost
momenta of typically  $P/M \ge 1$.  Only for very small boost momenta
P simple approximations, such as 
%
\renewcommand{\theequation}{16a}
\begin{equation}
u (z, \mbox{\boldmath$\rho$} , a) \cong u \left ( r, a/ \left (
\sqrt{3} \; \lambda (P) \right) \right) \quad \mbox{and} 
\end{equation}
%
%
\renewcommand{\theequation}{16b}
\begin{equation}
u( z, \mbox{\boldmath$\rho$} , R) = \exp ( - (r / (\sqrt{3} \; \lambda
(P) R)) ^{2} \, u (r, R) 
\end{equation}
\renewcommand{\theequation}{\arabic{equation}}
\setcounter{equation}{16}
simulate very qualitatively Lorentz quenching of slowly moving systems.
\end{itemize}
\vskip 0.2cm
With increasing boost momenta the breaking of the spherical symmetry
for the quenched bay leads for the ground (and all excited) state to
the admixture of additional angular momenta, which drastically enhance
the momentum spectrum of the ground state with increasing q. A
characteristic ressult is shown in Fig. 3 for the d-wave admixture for
different boost momenta.
\vskip 0.2cm
Summarizing our main findings in this note, we have formulated
covariant wave functions and their transformation properties in an
analytical quark - diquark model for the baryon and we find
characteristic modifications from the baryon rest system to moving
Lorentz-systems for different confining kernels.  
\vskip 0.2cm

Our findings suggest possible extensions and  basic shortcomings of
the model. We feel that an extension of the model to mesons as
$q\overline{q}$  systems, towards a more realistic quark-diquark
description of baryons or to genuine   3-quark systems (together with
a systematic inclusion of CM corrections) imposes only technical
problems and is certainly feasible. Here we only mention that the
quenching factor from eq.(7) is recovered in leading order for all
current projections of the BS equation: as an example the
Blankenbecler-Sugar reduction (30) for quarks with equal masses yields
immediately 
\begin{equation}
G_{BBS} \, (P, q) \sim \delta (q_{0} -  P/M  q_{z} )
\end{equation}
A more serious problem for confining kernels with a finite power in
the interquark distance r is the precise formulation of Lorentz
quenching for the kernels itself (in Bag models the dependence is
absorbed in the boundary condition). Here the unsurmountable problem
is our current lack in understanding the confining mechanism: it is
not clear, how the full P dependence enters into the kernel 
(for an example compare ref.(37); however, different
szenarios may lead to quantitatively very different results for large P;
for an example compare ref. (37)). We feel that a more realistic
extension of present phenomenological quark models undoubtedly
requires a much deeper analytical understanding of confinement. Here
significant
progress in various directions has been achieved recently, to mention
only the modelling of
confinement of QCD in the Coulomb gauge (38) or the  extension of
conecpt of instantons to merons as solutions of the classical QCD
equations (39).    

{\Large {\bf Figure Captions }}

\begin{enumerate}
\item[Fig. 1:]
Z-dependence of the large component of the harmonic oscillator ground state
in coordinate (a) and momentum space (b) for different boost momenta P
= 0 (thin line), M (middle line) and 2M (thick line).  

\item[Fig. 2:]
(a)
Quenching and boundary conditions for the bag along the boost momenta
P=0, M, 2\,M. The functions f(z) and g(z) denote the large and small
components of the ground state wave function (for a bag radius 
R = 1 fm). 
\vskip 0.2cm
 
\item[Fig. 3:]
D-state admixture to the harmonic oscillator ground state. Compared
are the s-wave distribution for P=0 with the d-state component for P=M
and 2\,M (middle and thick line, respectively). 

\end{enumerate}
\begin{figure}
\epsfig{file=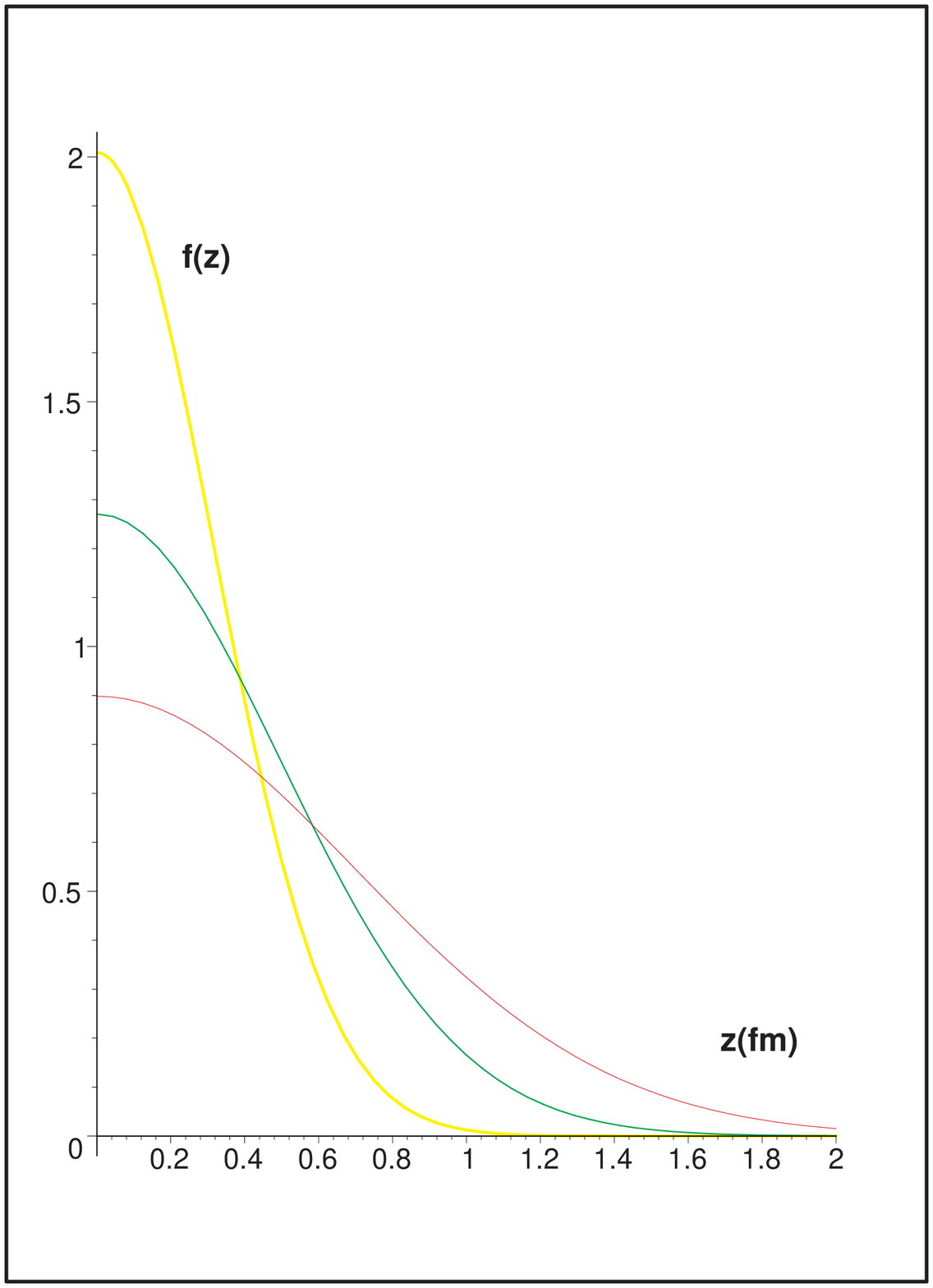}
\end{figure}
\begin{figure}
\epsfig{file=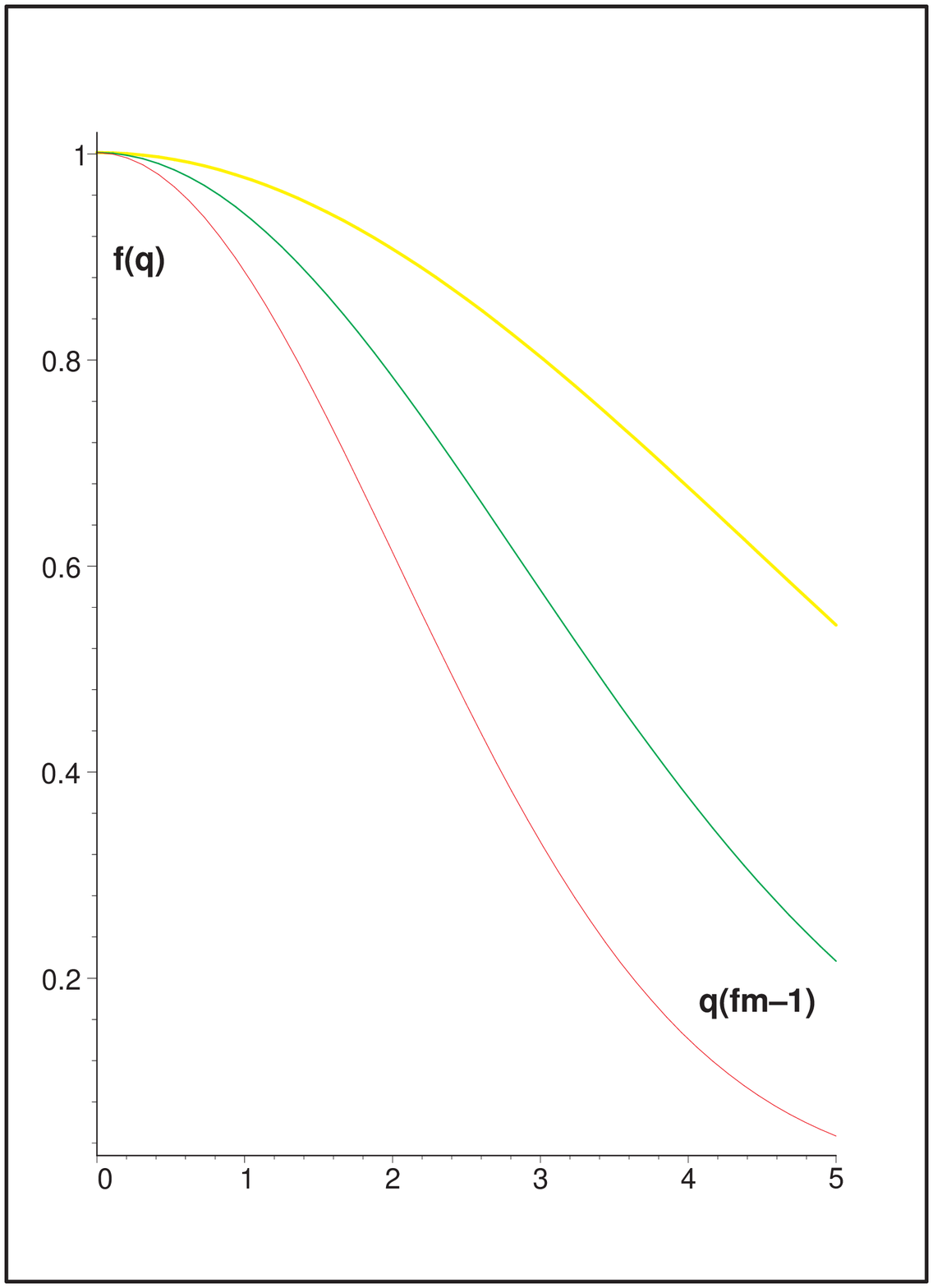}
\end{figure}
\begin{figure}
\epsfig{file=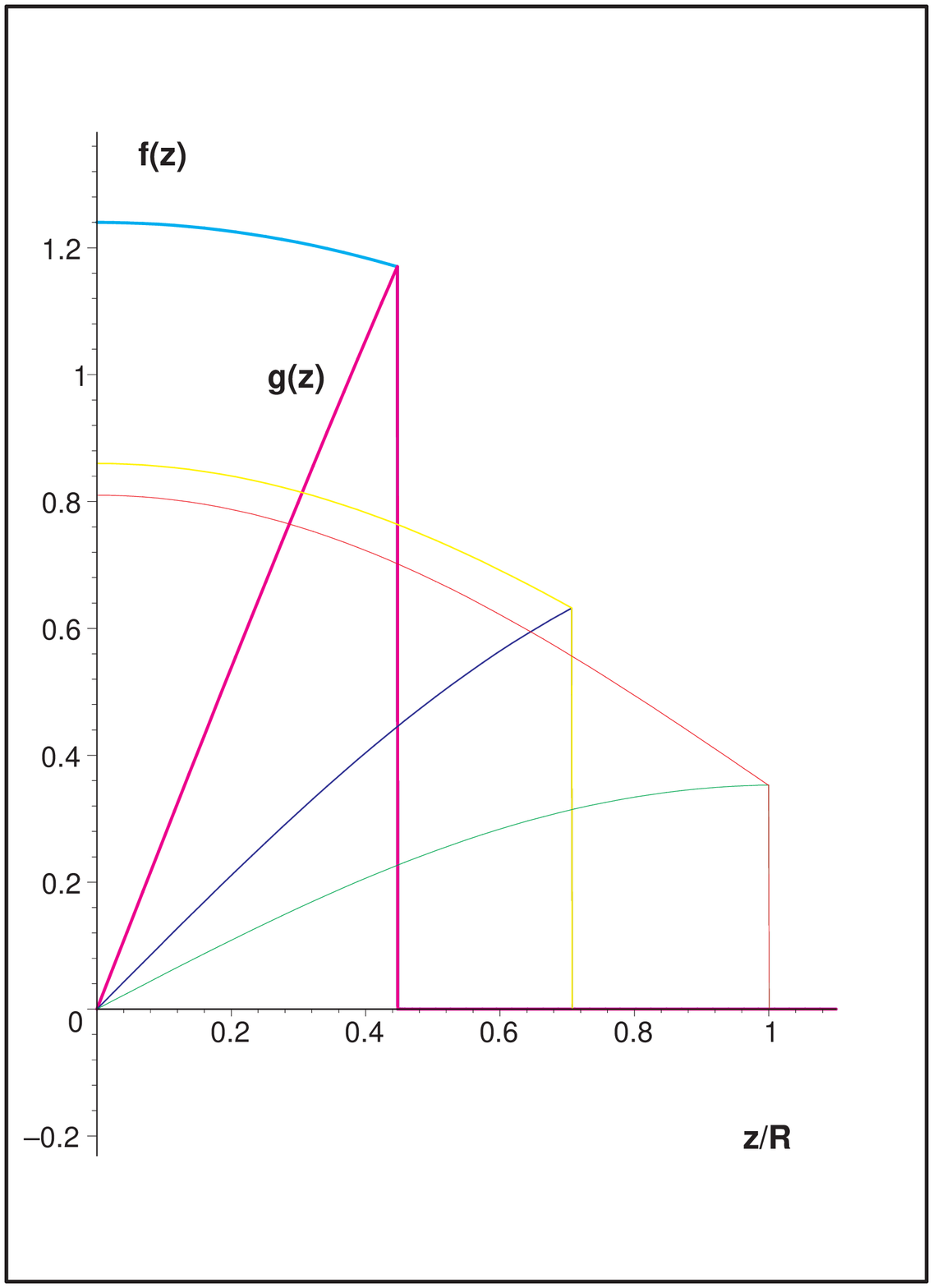}
\end{figure}
\begin{figure}
\epsfig{file=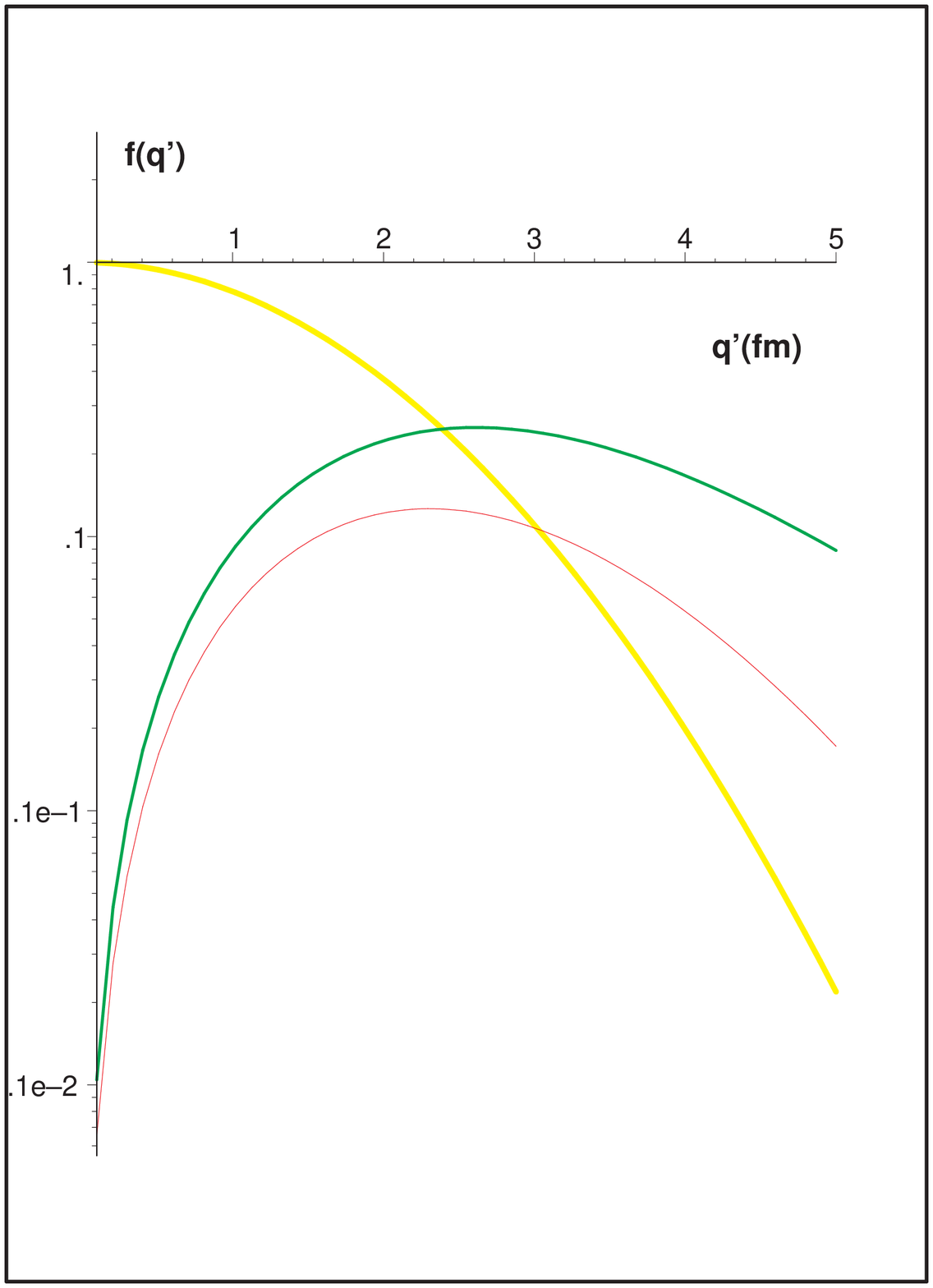}
\end{figure}

\end{document}